\documentclass[journal]{IEEEtran}
\IEEEoverridecommandlockouts

\usepackage{graphicx, subcaption}
\usepackage{amsmath,amsthm,amsfonts,amssymb}
\usepackage{cite}
\usepackage{bm}
\usepackage{url}
\usepackage{array}
\usepackage{color}
\usepackage{multirow}
\usepackage{booktabs,multirow}
\usepackage{xcolor}
\usepackage{enumerate}
\usepackage{enumitem}
\usepackage{makecell}
\usepackage{comment}

\usepackage[english]{babel}
\usepackage{algorithm}
\usepackage[noend]{algpseudocode}

\theoremstyle{plain}

\usepackage{mathtools}

\allowdisplaybreaks[4]

\DeclareMathOperator*{\argmax}{arg\,max}

\begin{document}

\addtolength{\textheight}{0.22in} % longer
\addtolength{\voffset}{-0.11in}

\addtolength{\textwidth}{0.22in} % wider
\addtolength{\hoffset}{-0.11in}

\title{{P-WRFGS: Pruning 3D Gaussians for Efficient Wireless Radiance Field Construction}}

\author{
Chenghong~Bian,~\IEEEmembership{Member,~IEEE},
Meng~Hua,~\IEEEmembership{Senior Member,~IEEE}, and
Deniz~G\"und\"uz,~\IEEEmembership{Fellow,~IEEE}
\thanks{
{C. Bian is with the Department of Electrical and Computer Engineering, Hong Kong University of Science and Technology.} {M. Hua} and D. G{\"u}nd{\"u}z are with the Department of Electrical and Electronic Engineering, Imperial College London (E-mail: m.hua@imperial.ac.uk). \textit{(Corresponding author: Meng Hua)}.
}\thanks{
This work received funding from the UKRI for the projects INFORMED-AI (EP/Y028732/1),  AI-R (ERC Consolidator Grant, EP/X030806/1) and SNS JU project 6G-GOALS under the EU’s Horizon program (Grant Agreement No. 101139232).
}

}

\maketitle

\begin{abstract}
Wireless channel modeling is a key building block for next-generation wireless systems. Predicting the channel state information (CSI) across different transmitter locations can substantially reduce the pilot and feedback overhead of conventional channel estimation. We propose {P-WRFGS}, an efficient wireless radiance field modeling framework built upon 3D Gaussian splatting. {P-WRFGS} introduces a learnable mask for each 3D Gaussian primitive to indicate its importance, which guides the pruning of less significant primitives for more efficient rendering. The model is trained using a weighted combination of rendering and regularization losses, allowing a flexible trade-off between rendering quality and efficiency. {Numerical results on the $\text{NeRF}^2$ dataset demonstrate that {P-WRFGS} achieves up to 100$\times$ storage reduction and 7$\times$ rendering speed-up with only mild degradation in SSIM and the achievable rate.} Moreover, initializing the Gaussian primitives from a 3D point cloud of the scene further improves the entire quality-efficiency trade-off.
\end{abstract}

\begin{IEEEkeywords}
Wireless channel modeling, 3D Gaussian splatting.
\end{IEEEkeywords}

\section{Introduction}\label{sec:intro}
Modern wireless networks use electromagnetic (EM) waves to deliver information over wireless channels. Owing to complex interactions (e.g., reflection, scattering, and diffraction) between EM waves and the environment, the channel state information (CSI) between a transmitter-receiver pair is hard to obtain in practice. The conventional approach is pilot-based estimation: the transmitter periodically transmits known pilot symbols, and the receiver estimates the channel from the received pilots and feeds back CSI to the transmitter \cite{channel_est, channel_fb}. However, this pilot-and-feedback based approach incurs significant power and communication overhead. It has therefore motivated the development of efficient channel modeling techniques that can provide accurate CSI without exhaustive measurements.

To tackle the complex wireless channel modeling problem, machine learning techniques have been proposed \cite{radio_map, learn_rt, 3d_radio_map}. In particular, the authors of \cite{radio_map} predict city-scale radio maps using a U-Net backbone that takes the city layout and transmitter positions as inputs. This approach performs well in scenarios with relatively few scatterers. To handle more complex propagation environments, the authors in \cite{learn_rt} adopt differentiable ray tracing, where each ray emitted from the transmitter is carefully traced through the environment until it reaches the receiver. The final received signal is computed by summing up the contributions of all candidate rays. Although accurate when material properties and geometry layout are perfectly known, ray tracing incurs substantial computation and latency, making it impractical for real-time CSI acquisition.

Recently, inspired by the success of learning-based 3D rendering methods such as neural radiance field (NeRF) \cite{nerf} and 3D Gaussian splatting (3DGS) \cite{3dgs}, {several works \cite{nerf2, wrfgspp, rf_3dgs, gs3d, point2radio} have proposed neural rendering frameworks for efficient channel modeling and signal recovery in wireless networks.} At a high level, 3D rendering aims to synthesize novel scenes from a limited set of captured images. This closely resembles the CSI acquisition task, which seeks to infer CSI at unobserved locations from CSI measurements at observed locations. Moreover, both visible light and wireless signals are EM waves and are governed by Maxwell's equations. Building on these similarities, following NeRF's volume-rendering paradigm, $\text{NeRF}^2$ \cite{nerf2} demonstrates strong CSI prediction performance. However, it requires dense sampling along each ray and neural network evaluation at every sampled point, which incurs substantial computational overhead. To address this limitation, WRF-GS \cite{wrfgspp} instead represents the wireless scene with a set of 3D Gaussian primitives whose parameters are functions of the transmitter position. Instead of sampling along the ray as in \cite{nerf2}, rasterization is employed to efficiently generate the CSI for a given transmitter position. Both WRF-GS \cite{wrfgspp} and its extended version \cite{urfgs} substantially improve the CSI prediction accuracy and achieve higher efficiency than prior approaches \cite{nerf2, learn_rt}. {GS3D~\cite{gs3d} combined 3DGS with stable diffusion for robust signal recovery in 3D wireless networks, highlighting the broader potential of Gaussian-based representations in wireless applications.}

In this paper, we extend the WRF-GS framework \cite{wrfgspp} to enable more efficient CSI prediction.
Recent work on efficient 3D rendering \cite{light_gaussian, compact_gaussian} observed that Gaussian primitives generated by 3DGS \cite{3dgs} are highly redundant, and pruning them reduces both storage and inference cost. In particular, \cite{light_gaussian, compact_gaussian} remove less important Gaussian primitives and replace spherical harmonics with multi-layer perceptrons (MLPs). While these techniques have been shown effective for novel scene synthesis, their applicability to the wireless domain remains unclear. Motivated by \cite{light_gaussian}, we propose {P-WRFGS} to enhance the efficiency of WRF-GS for real-time CSI acquisition. We associate a learnable mask with each Gaussian and render only those whose mask value exceeds a predefined threshold. The mask is binarized by a straight-through estimator so that the threshold is differentiable end-to-end.
The mask is jointly optimized with the {Gaussian primitives} using a loss function that combines CSI prediction error and the average mask value. Furthermore, we initialize the Gaussian centers from a 3D point cloud of the scene rather than uniformly at random. {Numerical results on the $\text{NeRF}^2$ dataset show that {P-WRFGS} achieves up to 100$\times$ storage reduction and 7$\times$ rendering speedup while incurring only mild degradation in SSIM and the achievable rate ratio with respect to the ground truth spatial spectrum.} We further provide ablation studies on the regularization weight, initialization strategy, and per-prediction latency to quantify the rendering-quality and efficiency trade-off.

\begin{figure*}[t]
  \centering
  \includegraphics[width=\linewidth]{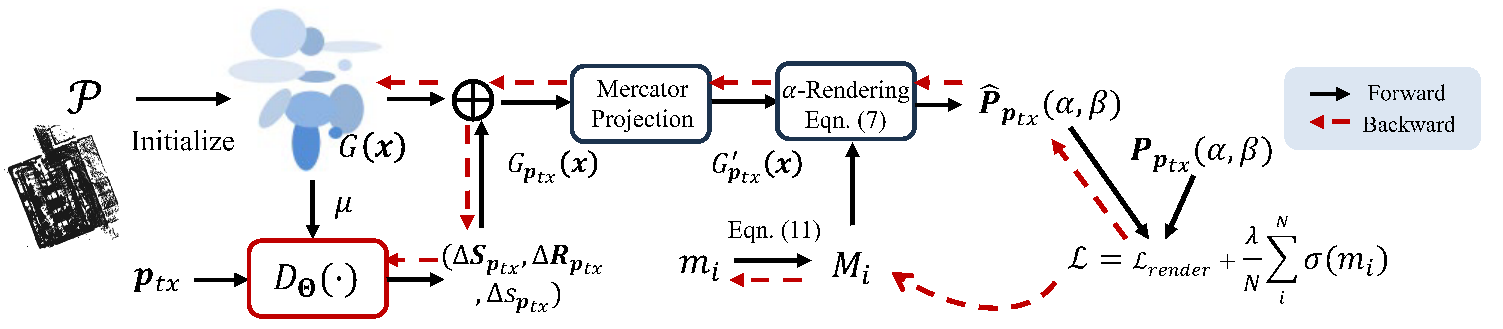}\\
  \caption{The flowchart of the proposed {P-WRFGS} framework. The 3D Gaussian primitives are initialized using the geometry layout of the scene, $\mathcal{P}$. A deformation block, $D_{\Theta}$, takes the transmitter location, $\bm{p}_{\mathrm{tx}}$ as input to achieve $\bm{p}_{\mathrm{tx}}$-dependent rendering. The masking operation is performed for efficient CSI prediction. Finally, a loss function combining rendering and regularization terms is calculated to update all learnable parameters.}
\label{fig:system_model}
\end{figure*}

\section{System Model}\label{sec:II}
A generic wireless communication system consists of a transmitter that modulates information bits onto wireless signals, propagated through the channel. Due to reflection, scattering, and diffraction, the received signal at the receiver (assuming single receive antenna) is a superposition of multiple copies of the transmitted signal, which can be considered as virtual transmitters:
\begin{equation}
y = A e^{j \phi} \sum_{l=0}^{L-1} \Delta A_l e^{j \Delta \phi_l},
\label{eq:multipath}
\end{equation}
where $A$ and $\phi$ denote the amplitude and phase of the transmitted signal, while $\Delta A_l$ and $\Delta \phi_l$ represent the attenuation and phase shift of the $l$-th propagation path at the receiver, $l \in \{0,\ldots,L-1\}$, respectively.

We consider a 2D uniform planar array of $K$ antennas, with $\sqrt{K}$ antennas per dimension. We are interested in the angle-of-arrival (AoA) of the received signal, which can be estimated using the phases of the received signals across antennas.
In particular, we define the spatial spectrum $\bm{P}=\{P(\alpha,\beta)\}$, which, for a given azimuth--elevation pair $(\alpha,\beta), \alpha\in [0, 2\pi], \beta\in [0, \frac{\pi}{2}]$, is calculated as
\begin{equation}
P(\alpha,\beta)=
\frac{1}{K}
\left|
\sum_{m,n=0}^{\sqrt{K}-1}
e^{j \big(\Delta \hat{\theta}_{m,n}-\Delta \theta_{m,n}\big)}
\right|^{2},
\label{eq:spatial_spectrum}
\end{equation}
where $\Delta \hat{\theta}_{m,n} = \hat{\theta}_{m,n}-\hat{\theta}_{0,0}$ is the measured phase difference between antenna $(m,n)$ and the reference antenna $(0,0)$, and $\Delta \theta_{m,n}$ is the corresponding theoretical phase difference for a plane wave from direction $(\alpha, \beta)$. As shown in \cite{nerf2, wrfgspp}, $P(\alpha,\beta)$ represents the angular power distribution, thus characterizing the wireless radiation field. With an angular resolution of $1^{\circ}$ for both azimuth and elevation angles, the spatial spectrum can be organized into a $360\times90$-pixel image.

Throughout this paper, the location and orientation of the receiver equipped with the aforementioned 2D antenna array are fixed, while the transmitter location, denoted as $\bm{p}_{\mathrm{tx}}$, varies within the scene. Consequently, the spatial spectrum depends on $\bm{p}_{\mathrm{tx}}\in \mathbb{R}^3$, and our goal is to predict $P_{\bm{p}_{\mathrm{tx}}}(\alpha,\beta)$ for different transmitter positions.
To evaluate the prediction accuracy, we adopt the structural similarity index measure (SSIM), widely used in computer vision \cite{nerf}.

\section{The {P-WRFGS} Scheme}\label{sec:III}
In this section, we first recap the WRF-GS scheme proposed in \cite{wrfgspp}. Then, we present the proposed {P-WRFGS} framework, which reduces the number of 3D Gaussian primitives for lower storage cost and improved rendering efficiency. Finally, the training procedure of {P-WRFGS} is presented.

\subsection{WRF-GS Recap}
To predict the spectrum $P_{\bm{p}_{\mathrm{tx}}}(\alpha, \beta)$ for the transmitter location $\bm{p}_{\mathrm{tx}}$, WRF-GS \cite{wrfgspp} employs 3D Gaussian primitives to represent each of the virtual transmitters \cite{3dgs}. Each Gaussian primitive is parameterized by the center position $\bm{\mu} \in \mathbb{R}^{3}$, the covariance matrix $\bm{\Sigma} \in \mathbb{R}^{3 \times 3}$, the opacity $o \in (0, 1)$, and the corresponding complex signal $s \in \mathbb{C}^1$. Similarly to \cite{3dgs}, the 3D Gaussian distribution is defined as
\begin{equation}
G(\bm{x}) = \exp \!\left(-\frac{1}{2} (\bm{x}-\boldsymbol{\mu})^{\top} \boldsymbol{\Sigma}^{-1} (\bm{x}-\boldsymbol{\mu}) \right),
\label{eq:gaussian_3d}
\end{equation}
and the covariance matrix can be factorized into a rotation matrix ($\bm{R}$) and a scaling matrix ($\bm{S}$) as
\begin{equation}
\boldsymbol{\Sigma} = \bm{R} \bm{S} \bm{S}^{\top} \bm{R}^{\top}.
\label{eq:covariance}
\end{equation}

The rendering output, i.e., the spatial spectrum, should be a function of $\bm{p}_{\mathrm{tx}}$ and additional processing is required.
Following \cite{wrfgspp}, we employ a deformation block, denoted by $D_{\Theta}$, which takes the center position of the $i$-th Gaussian, $\bm{\mu}_i$, and the transmitter location $\bm{p}_{\mathrm{tx}}$ as inputs. It outputs the corresponding calibration terms, $\Delta \bm{S}_i(\bm{p}_{\mathrm{tx}}), \Delta \bm{R}_i(\bm{p}_{\mathrm{tx}}), \Delta s_i(\bm{p}_{\mathrm{tx}})$, for the scaling matrix $\bm{S}_i$, the rotation matrix $\bm{R}_i$, and the signal $s_i$, respectively:
\begin{equation}
{D}_{\Theta} : \big(\bm{\mu}_i, \bm{p}_{\mathrm{tx}}\big)
\;\Rightarrow\;
\big(\Delta \bm{S}_i(\bm{p}_{\mathrm{tx}}), \Delta \bm{R}_i(\bm{p}_{\mathrm{tx}}), \Delta s_i(\bm{p}_{\mathrm{tx}})\big).
\label{eq:dtheta}
\end{equation}
Positional embedding is also adopted to provide the deformation block with richer features, which we refer to \cite{wrfgspp} for more details.
These calibration terms are added to the base parameters to form the $\bm{p}_{\mathrm{tx}}$-dependent Gaussian primitives, denoted as $G_{\bm{p}_{\mathrm{tx}}}(\bm{x}_i)$, for the subsequent rendering process:
\begin{equation}
    s_{i}(\bm{p}_{\mathrm{tx}}) = s_i + \Delta s_i(\bm{p}_{\mathrm{tx}}),
    \label{eq:cali}
\end{equation}
and the scaling and rotation matrices, $\bm{S}_{i}(\bm{p}_{\mathrm{tx}})$ and $\bm{R}_{i}(\bm{p}_{\mathrm{tx}})$, can be obtained similarly by replacing $s_i$ in \eqref{eq:cali} with $\bm{S}_i$ and $\bm{R}_i$.

After calibration, each 3D Gaussian, denoted by $G_{i,\bm{p}_{\mathrm{tx}}}(\bm{x})$, is projected following the Mercator projection proposed in \cite{wrfgspp}, resulting in a 2D Gaussian $G'_{i,\bm{p}_{\mathrm{tx}}}(\bm{x}')$ in the screen space. These projected Gaussians are sorted by their depth and rasterized in parallel across image tiles.
In particular, for a pixel at position $\bm{z}_s = (\alpha_s, \beta_s)$, its predicted spectrum value is computed through $\alpha$-blending over the $N_G$ Gaussians that actively contribute to the pixel value:
\begin{align}
\hat{P}_{\bm{p}_{\mathrm{tx}}}(\alpha_s, \beta_s) &=
\sum_{i=1}^{N_G}
s_{i}(\bm{p}_{\mathrm{tx}})\gamma_{i}(\bm{p}_{\mathrm{tx}})
\prod_{j=1}^{i-1}(1-\gamma_{j}(\bm{p}_{\mathrm{tx}})),
\label{eq:alpha_blend}
\\
\gamma_{i}(\bm{p}_{\mathrm{tx}}) &= o_{i} \, G'_{i, \bm{p}_{\mathrm{tx}}}(\Delta \bm{z}_{i}),
\end{align}
where $\hat{P}_{\bm{p}_{\mathrm{tx}}}$ denotes the predicted $\bm{p}_{\mathrm{tx}}$-dependent spatial spectrum, and $\Delta \bm{z}_{i} = \bm{z}_{i}-\bm{z}_{s}$ is the offset between the projected 2D Gaussian center and the pixel position. The resulting spectrum is compared with the ground-truth spectrum, ${\bm{P}}_{\bm{p}_{\mathrm{tx}}}$, to compute the rendering loss \cite{3dgs, wrfgspp} defined as:
\begin{equation}
    \mathcal{L}_{\text{render}} = (1-w)\mathcal{L}_1 + w\mathcal{L}_{\text{D-SSIM}},
    \label{eq:standard_loss}
\end{equation}
where $\mathcal{L}_1$ denotes the $\ell_1$-loss between the two spectra and $\mathcal{L}_{\text{D-SSIM}} \triangleq 1 - \text{SSIM}(\hat{\bm{P}}_{\bm{p}_{\mathrm{tx}}}, {\bm{P}}_{\bm{p}_{\mathrm{tx}}})$. The weight is set to $w = 0.2$ as in \cite{wrfgspp}.
The loss is backpropagated to update the {Gaussian primitives} and the deformation block.

\subsection{Reducing Learnable Parameters}\label{sec:IIIA}

We next introduce the methodology to reduce the number of Gaussians, thereby lowering storage and rendering costs. To achieve this, we assign a learnable mask score, denoted $m_i\in \mathbb{R}, i\in [1, N]$, to each 3D Gaussian primitive $G(\bm{x}_i)$. The sigmoid value $\sigma(m_i)$ acts as the importance score of the primitive to represent the wireless radiance field, and primitives with higher scores are more likely to be retained.
To facilitate the rasterization process, a binary mask is needed. We obtain $M_i \in \{0, 1\}$ by binarizing $\sigma(m_i)$ using a predetermined threshold of 0.01:
\begin{equation}
    M_i =
    \begin{cases}
        1, & \text{if } \sigma(m_i) \ge 0.01, \\
        0, & \text{if } \sigma(m_i) < 0.01.
    \end{cases}
    \label{eq:binary}
\end{equation}
During rendering, these masks are directly multiplied with the opacity, $o_i$, and the scaling matrix, $\bm{S}_i(\bm{p}_{\mathrm{tx}})$, to generate the spatial spectrum. Thus, Gaussian primitives with $\sigma(m_i)<0.01$ are skipped, accelerating the spectrum prediction process.

However, \eqref{eq:binary} is non-differentiable. Thus, we adopt the straight-through estimator (STE) during training.
In particular, we adopt
\begin{align}
    M_i = \operatorname{sg}\!\big( \mathbf{1}[\sigma(m_i) \ge 0.01] - \sigma(m_i) \big) + \sigma(m_i), \label{eq:mask}
\end{align}
where $\mathbf{1}[\cdot]$ is the indicator function and $\operatorname{sg}(\cdot)$ stands for the stop-gradient operator.
This formulation enables efficient rendering in the forward pass thanks to the binarized masks, while allowing gradients to update the mask score $m_i$ during backpropagation.

To strike a balance between the rendering quality and efficiency, we introduce a regularization term $\mathcal{L}_m$:
\begin{equation}
    \mathcal{L}_m = \frac{1}{N} \sum_{i=1}^{N} \sigma(m_i), \label{eq:mask_loss}
\end{equation}
{{where $N$ denotes the number of Gaussian primitives.}} This term corresponds to the expected ratio of retained Gaussians.
The overall loss function can be expressed as the weighted combination of the rendering loss $\mathcal{L}_{\text{render}}$ introduced in \eqref{eq:standard_loss} and the regularization loss:
\begin{equation}
    \mathcal{L} = \mathcal{L}_{\text{render}} + \lambda \mathcal{L}_m,
    \label{eq:total_loss}
\end{equation}
where $\lambda$ balances the quality and efficiency trade-off.
During training, Gaussians with $M_i = 0$ are discarded every $I_p$ iterations in the pruning phase, as detailed in the subsequent subsection.

\begin{algorithm}[t!]
\caption{Training of {P-WRFGS} with Learnable Masks.}\label{alg:eff_wrfgs}
\begin{algorithmic}[1]
\State \textbf{Input:} $\mathcal{P}=\{\bm{x}_i\}_{i=1}^{|\mathcal{P}|}$, $M, M_d, M_p$, $I_p$, $\epsilon$, $\lambda, \{(\bm{p}_{\mathrm{tx}},\bm{P}_{\bm{p}_{\mathrm{tx}}})\}$
\State \textbf{Output:} $\{\bm{\mu}_i,\bm{\Sigma}_i, o_i, s_i\}_{i=1}^{N}$, $D_\Theta$
\State \textbf{Initialization from $\mathcal{P}$: $\bm{\mu}_i \leftarrow \bm{x}_i, \bm{\Sigma}_i \leftarrow d_i^2\bm{I}_3$} where $\bm{x}_i \in \mathcal{P}, d_i \leftarrow$ Eqn. \eqref{eq:di}.

\For{$t=1$ to $M$}
\State Compute $M_i$ according to \eqref{eq:mask}.
\State Calculate $\bm{p}_{\mathrm{tx}}$-dependent 2D Gaussian, $G'_{i, \bm{p}_{\mathrm{tx}}}$.
\State Render $\bm{P}_{\bm{p}_{\mathrm{tx}}}$ via Eqn. \eqref{eq:alpha_blend} with $\gamma_i$ replaced by $M_i \gamma_i$.
\State Calculate $\mathcal{L}_{\text{render}}$ and $\mathcal{L}_m$ as per Eqn. \eqref{eq:standard_loss}, \eqref{eq:mask_loss}.
\State \textbf{if} $t \le M_d$ \textbf{then} Minimize $\mathcal{L} = \mathcal{L}_{\text{render}} + \lambda \mathcal{L}_m$ and
\parbox[t]{0.82\linewidth}{%
\hspace*{2em}perform Gaussian densification \cite{3dgs, wrfgspp}.}
\State \textbf{else if} $M_d < t \le M_p$ \textbf{then} Minimize $\mathcal{L}$ and prune
\parbox[t]{0.82\linewidth}{%
\hspace*{2em}Gaussians with $M_i=0$ if $t \bmod I_p = 0$.}
\State \textbf{else} Minimize $\mathcal{L}_{\text{render}}$.
\EndFor
\end{algorithmic}
\end{algorithm}

\subsection{Training Methodology}

\subsubsection{Initialization of 3D Gaussian Primitives}
The authors of \cite{wrfgspp} use random initialization.  We instead initialize from the 3D point cloud of the scene provided in \cite{nerf2}, which we find improves the quality–efficiency trade-off (Fig. \ref{fig:eval_eff_wrfgs_b}), since reflection, scattering, and diffraction occur on object surfaces. Incorporating the geometric layout helps the model infer the positions of virtual transmitters, i.e., the centers of the 3D Gaussian primitives. Thus, we set the number of initialized Gaussian primitives equal to $|\mathcal{P}|$, i.e., the cardinality of the point cloud. For the $i$-th Gaussian primitive, we set $\bm{\mu}_i = \bm{x}_i$ and $\bm{\Sigma}_i = d_i^2 \bm{I}_3$, where $d_i$ denotes the $\ell_2$-distance between the center of the $i$-th Gaussian and its nearest neighbor:
\begin{equation}
    d_i \triangleq\min_{j\neq i} \|\bm{\mu}_i - \bm{\mu}_j\|_2.
    \label{eq:di}
\end{equation}

\subsubsection{Density Control Procedure}
After initialization, we optimize the parameters of the 3D Gaussian primitives and the deformation network $D_{\Theta}$ using the loss functions defined in \eqref{eq:standard_loss} and \eqref{eq:total_loss}. The entire training phase is partitioned into the densification, pruning, and fine-tuning phases, as detailed next.

Training starts with $M_d$ iterations of the densification phase. Following \cite{wrfgspp}, we only densify the Gaussians in this phase to ensure a sufficient representation capacity of the scene. In this phase, the model is trained using \eqref{eq:total_loss}. In the pruning phase, which lasts for $M_p-M_d$ iterations, we stop densification and prune Gaussian primitives every $I_p$ iterations according to the learnable masks. The number of Gaussians, $N$, is reduced after each pruning operation. Since the remaining Gaussian primitives satisfy $\sigma(m_i)\ge 0.01$, the regularization term $\mathcal{L}_m$ increases immediately after pruning. The model is then optimized to reach a new equilibrium between rendering efficiency and quality by reducing the average mask scores, leading to a steadily decreasing Gaussian count.
Finally, in the fine-tuning phase, we stop pruning the Gaussian primitives and only optimize the rendering quality. 
%Thus, we adopt the loss function defined in \eqref{eq:standard_loss}, and the masks are not updated. 
The parameters of the 3D Gaussian primitives and the weights of the deformation block, $\Theta$, are optimized for $M - M_p$ additional iterations to improve the rendering performance. {We note that the training schedule illustrated above mitigates premature deletion of important Gaussian primitives by first forming an overcomplete representation, pruning only every \(I_p\) iterations, and finally fine-tuning without pruning. Experiments verified the robustness of the {P-WRFGS} models when trained with multiple random seeds.} The overall optimization process is summarized in Algorithm \ref{alg:eff_wrfgs}.

\section{Numerical Experiments}\label{sec:IV}

\begin{figure}[t]
     \centering
     \begin{subfigure}{0.95\columnwidth}
         \centering
         \includegraphics[width=\columnwidth]{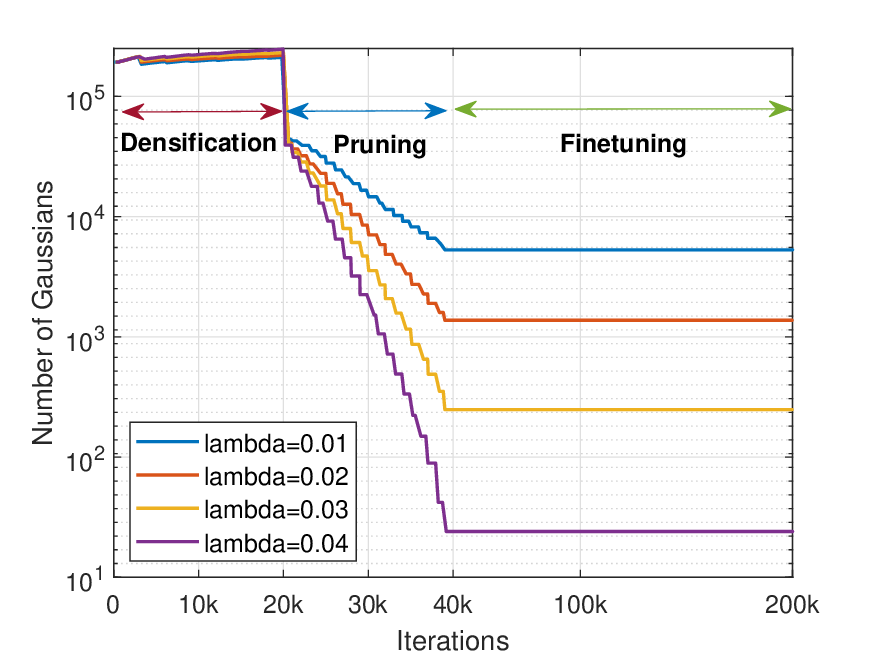}
         \caption{}
         \label{fig:eval_eff_wrfgs_a}
     \end{subfigure}
     \vspace{1cm}
     \begin{subfigure}{0.95\columnwidth}
         \centering
         \includegraphics[width=\columnwidth]{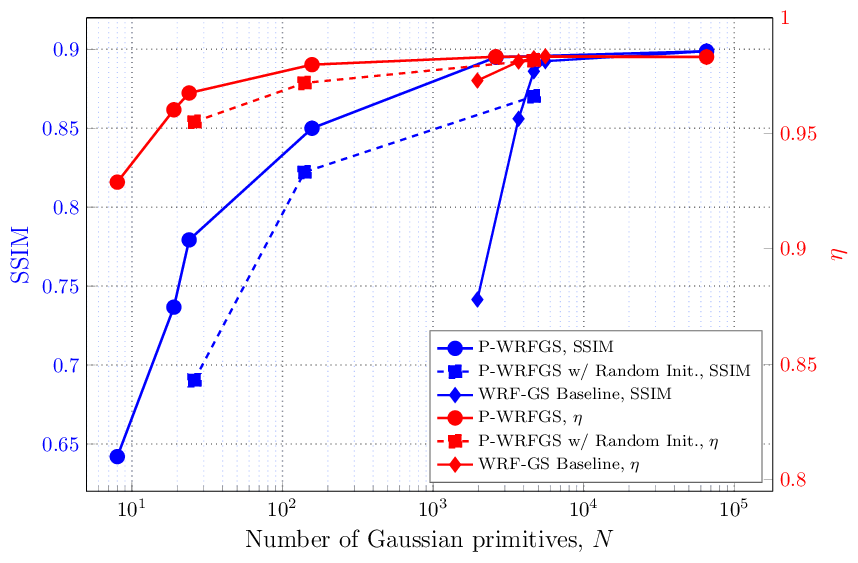}
         \caption{}
         \label{fig:eval_eff_wrfgs_b}
     \end{subfigure}
     \vspace{-0.5cm}
  \caption{{Performance evaluation of the {P-WRFGS} scheme: (a) The evolution of the number of Gaussian primitives, $N$, w.r.t. the optimization iterations. (b) The SSIM and $\eta$ trade-offs versus $N$ for {P-WRFGS} w/, w/o point cloud initialization, and the WRF-GS baseline.}}
  \label{fig:eval_eff_wrfgs}
\end{figure}

We evaluate the proposed {P-WRFGS} framework on a real-measurement dataset in \cite{nerf2}. The dataset consists of 4898 training and 1225 test samples, where each sample comprises a spatial spectrum and the corresponding transmitter location, $\bm{p}_{\mathrm{tx}}$. We adopt $M = 2\times 10^5$ training iterations and set $M_d = 2\times10^4$, and $M_p = 4\times 10^4$ for the densification and pruning phases. {During the pruning phase, Gaussians with $\sigma(m_i) < 0.01$ and opacity $o_i < \epsilon = 0.005$ are pruned every $I_p = 1000$ iterations\footnote{{
We further note that moderate changes in \(\epsilon\) and \(I_p\) do not collapse the efficiency and accuracy trade-off: a smaller \(\epsilon\) or larger \(I_p\) slightly improves system performance at the cost of more Gaussian primitives.}}.} Due to the page limit, we report only the spatial-spectrum results. %{P-WRFGS} extends naturally to received signal strength indicator (RSSI) and full CSI prediction.

\subsection{Efficiency and Accuracy Trade-off}
{Fig. \ref{fig:eval_eff_wrfgs}(a) shows the evolution of the number of Gaussian primitives, $N$, during training.} We train one model per $\lambda \in \{0.01, 0.02, 0.03, 0.04\}$ and initialize them using the point cloud provided in \cite{nerf2}, which consists of roughly $1.9\times 10^5$ points.
As can be seen, the number of Gaussians increases during the densification phase, as the wireless radiance field is being built. During the pruning phase, the number decreases every $I_p$ iterations because some Gaussian primitives are pruned according to the learnable masks. For larger $\lambda$, the count drops more rapidly and converges to a smaller value. In the fine-tuning phase, $N$ is held constant, and only the {Gaussian primitives} and the deformation network are updated to improve rendering quality.

{
Fig.~\ref{fig:eval_eff_wrfgs}(b) compares {P-WRFGS} with and without point-cloud initialization  and a WRF-GS baseline with higher pruning thresholds, $\epsilon\in\{0.005,0.1,0.2,0.4,0.5\}$. 
As shown in the figure, the proposed {P-WRFGS} achieves the best efficiency and accuracy trade-off among the schemes considered. 
%For point-cloud initialized {P-WRFGS}, we use $\lambda\in\{0,0.01,0.02,0.03,0.035,0.04\}$, where $\lambda=0$ corresponds to WRF-GS~\cite{wrfgspp} with $\epsilon=0.005$. As $\lambda$ increases, the learned masks progressively reduce the number of Gaussian primitives. Specifically, $N$ decreases from $65469$ at $\lambda=0$ to $157$ at $\lambda=0.02$, with a mild degradation in the SSIM performance from $0.899$ to $0.850$. We also observe that only $8$ Gaussians are preserved under more aggressive pruning where $\lambda=0.04$. In comparision, 
In particular, the randomly initialized {P-WRFGS} scheme is slightly inferior to the proposed {P-WRFGS} scheme due to the lack of prior geometric information, yet the SSIM of the WRF-GS baseline drops significantly when the pruning threshold, $\epsilon$, increases. This is expected as increasing $\epsilon$ may remove important primitives with small opacity values, thus leading to sub-optimal performance.

We further evaluate the achievable-rate ratio to assess the potential of the proposed {P-WRFGS} model when applied to practical communication systems. In particular, let $\hat{P}(\alpha,\beta)$ and $P(\alpha,\beta)$ denote the predicted and ground-truth spatial-spectrum, respectively, at azimuth--elevation direction $(\alpha,\beta)$. The beam direction selected from the predicted spectrum is
$(\hat{\alpha},\hat{\beta})
=
\underset{\alpha,\beta}{\operatorname*{arg\,max}}\,
\hat{P}(\alpha,\beta)$,
while the optimal direction under the ground-truth spectrum is
$(\alpha^\star,\beta^\star)
=
\underset{\alpha,\beta}{\operatorname*{arg\,max}}\,
P(\alpha,\beta)$.
The achievable-rate ratio is defined as
\begin{equation}
\eta(\rho)
=
\frac{
\log_2\!\left(1+\rho P(\hat{\alpha},\hat{\beta})\right)
}{
\log_2\!\left(1+\rho P(\alpha^\star,\beta^\star)\right)
},
\label{eq:eta_revised}
\end{equation}
where $\rho$ denotes the transmit SNR, and we report $\eta(1)$, which corresponds to a transmit SNR of $0$ dB.
As shown in the same figure, it is surprising to see that when $\lambda=0.04$, where only $N=8$ Gaussians remain and the SSIM decreases to $0.642$, the achievable-rate ratio remains $0.9288$.  This suggests that the proposed framework can retain a high fraction of the optimal achievable rate even under extreme pruning, which is encouraging for practical applications. 
%In summary, the proposed {P-WRFGS} with point-cloud initialization consistently outperforms the remaining baselines in terms of both SSIM and achievable-rate ratio, demonstrating its effectiveness.
}

We further perform an ablation study to confirm that the rendering efficiency of the proposed scheme is greatly improved with a non-zero $\lambda$. All variants are evaluated on a server with NVIDIA RTX 1080 GPU. As shown in Table \ref{tab:latency}, $\lambda = 0$ baseline (which corresponds to WRF-GS with point-cloud initialization) is about 7$\times$ slower than the models with $\lambda = 0.01$ and $\lambda = 0.02$. {A latency floor is observed for the two pruned models, suggesting that the deformation MLP and per-frame projection dominate the total cost at this scale.} Table \ref{tab:latency} also lists the latency of the $\text{NeRF}^2$ \cite{nerf2} and ray-tracing baselines for context.

\begin{table}[tbp]
\caption{Per-prediction latency for {P-WRFGS} at three $\lambda$ values evaluated on RTX 1080 and for $\text{NeRF}^2$ and ray-tracing baselines.}
\begin{center}
\begin{tabular}{c|c|c|c|c|c}
\hline

\cline{1-6}
\textbf{Schemes} & \begin{tabular}{@{}c@{}}Proposed\\$\lambda = 0$ \end{tabular}   & \begin{tabular}{@{}c@{}}Proposed\\$\lambda = 0.01$ \end{tabular} & \begin{tabular}{@{}c@{}}Proposed\\$\lambda = 0.02$ \end{tabular} & $\text{NeRF}^2$ & \begin{tabular}{@{}c@{}}Ray\\Tracing \end{tabular}\\
\hline

\textit{$N$} & {65469} & {2613} & {157} & - & -\\
\hline
\textit{Latency} (s) & 0.22 & 0.033 & 0.031 & 0.2 & 7\\

\hline
\end{tabular}
\label{tab:latency}
\end{center}
\end{table}

\begin{figure}
     \centering
     \begin{subfigure}{0.32\columnwidth}
         \centering
         \includegraphics[width=0.8\columnwidth]{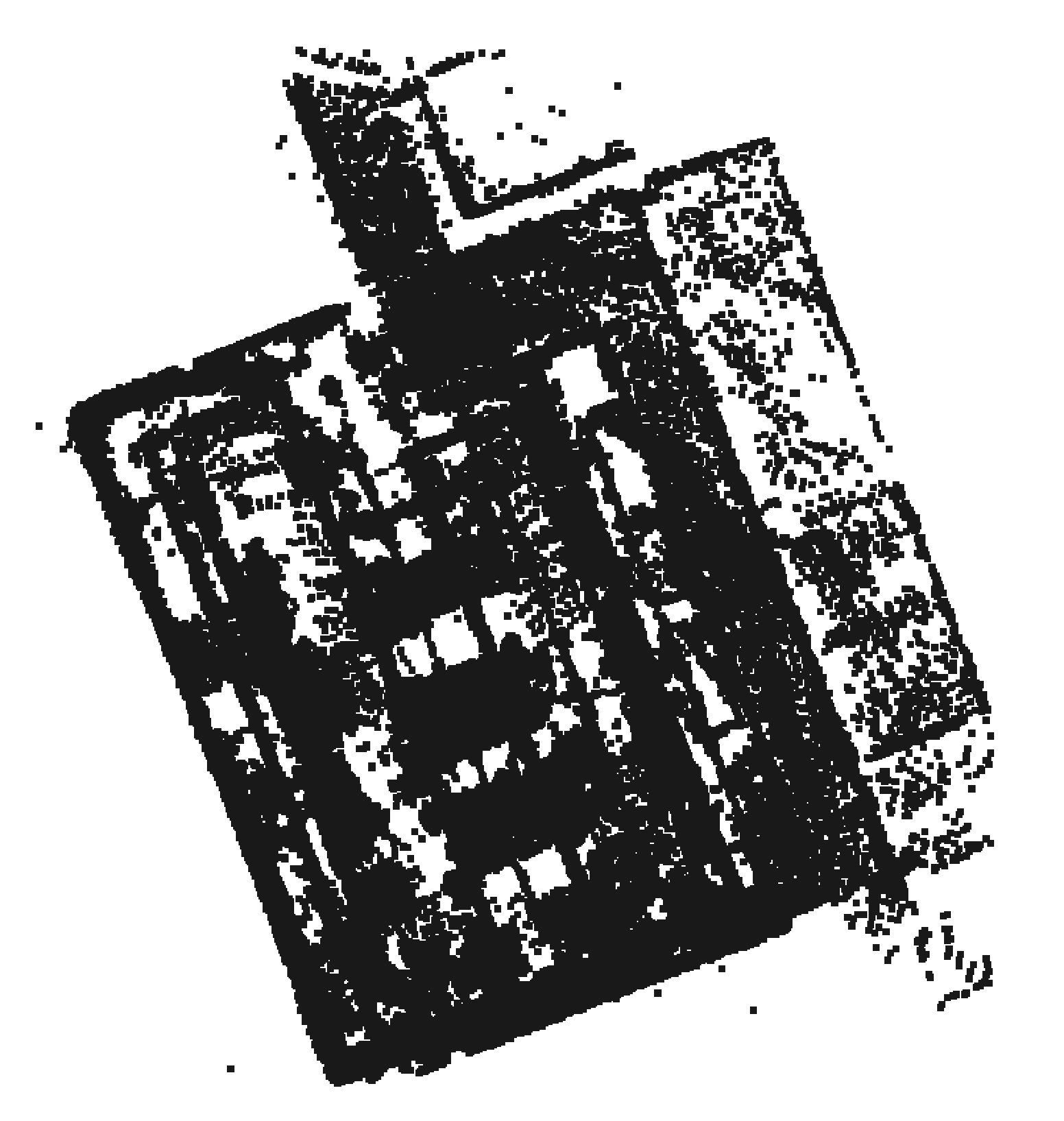}
     \end{subfigure}
     \begin{subfigure}{0.32\columnwidth}
         \centering
         \includegraphics[width=0.88\columnwidth]{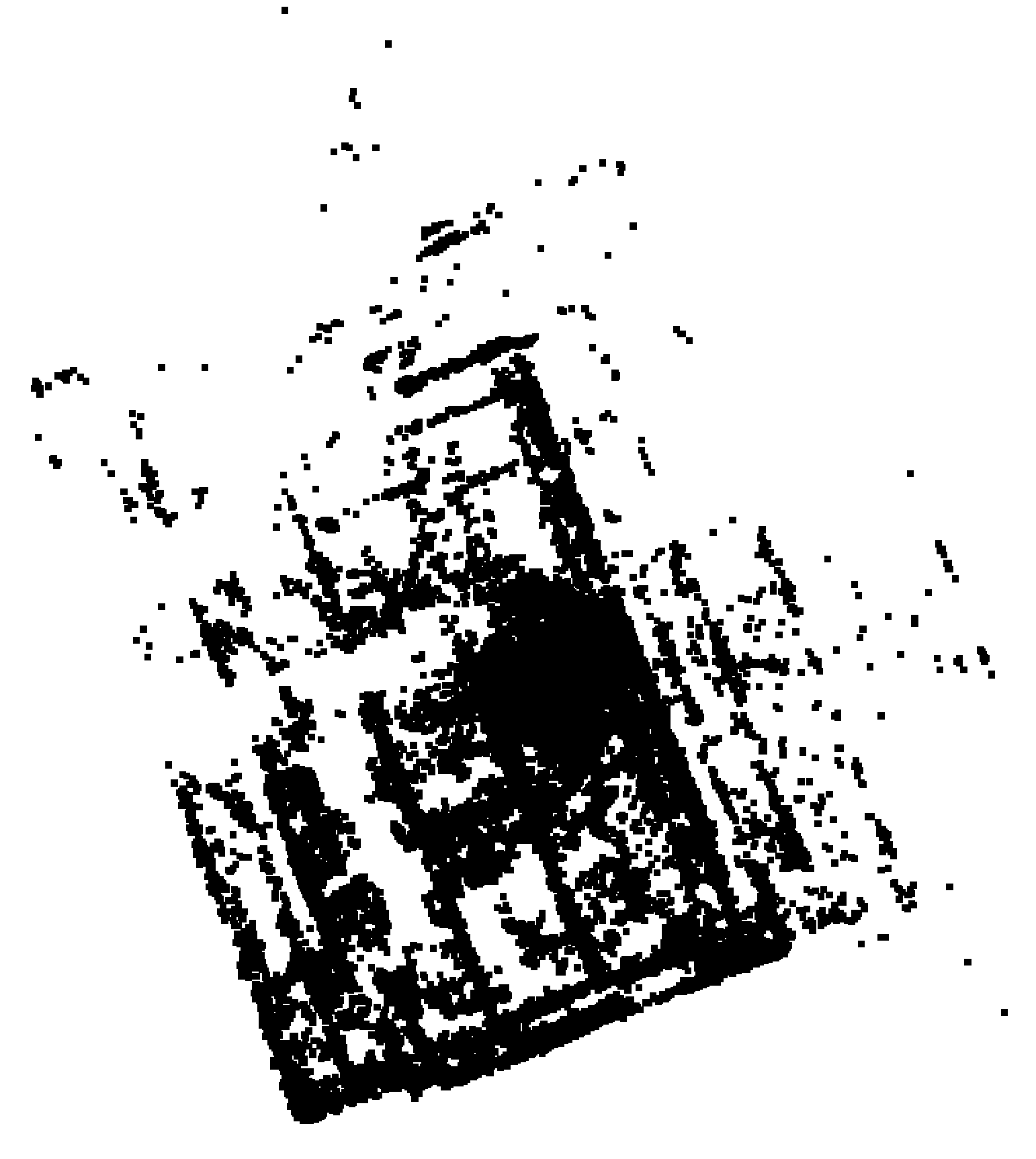}
     \end{subfigure}
     \begin{subfigure}{0.32\columnwidth}
         \centering
         \includegraphics[width=0.88\columnwidth]{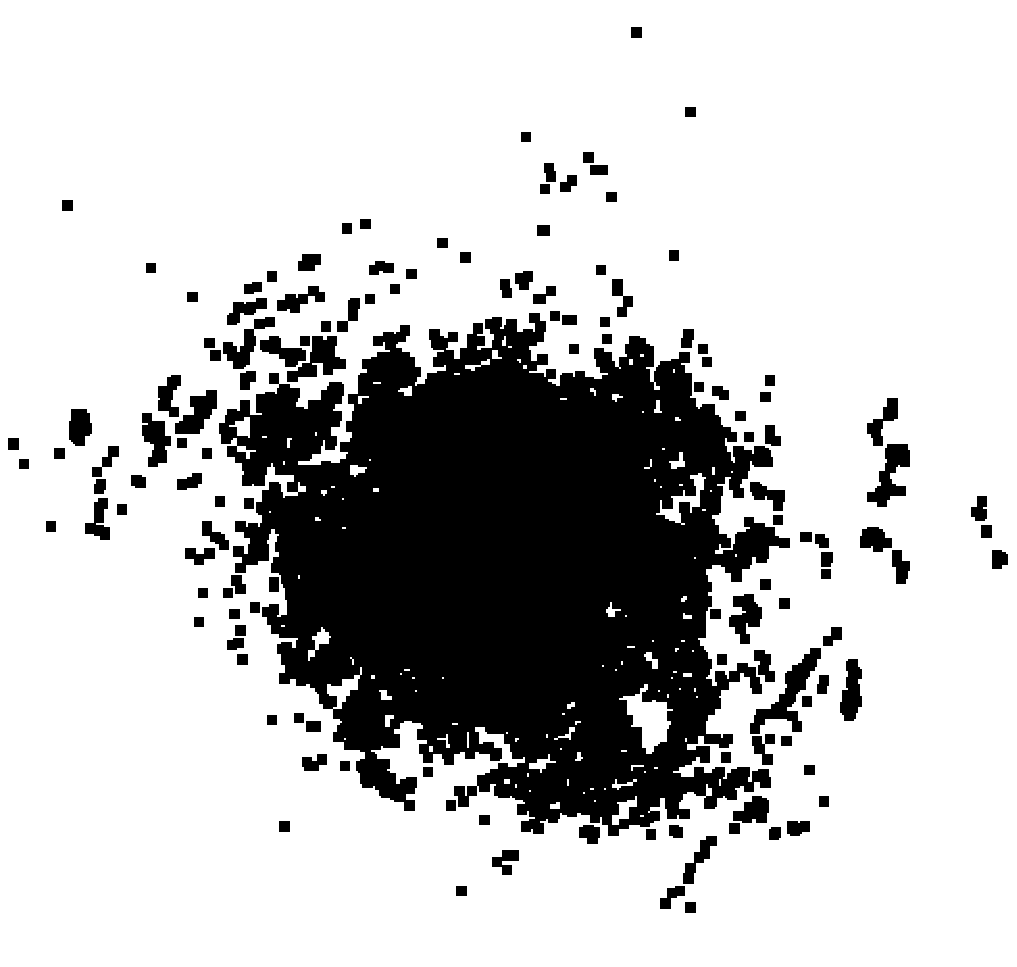}
     \end{subfigure}
  \caption{{Visualization of {P-WRFGS}. }Left: original $\mathcal{P}$; middle: 3D Gaussians initialized from $\mathcal{P}$; right: 3D Gaussians with random initialization.}
\label{fig:visualize_pc}
\end{figure}

\begin{figure}[t]
    \centering
    \captionsetup[subfigure]{justification=centering, labelformat=empty}
    \captionsetup{font={normalsize}}

    \begin{subfigure}[t]{0.19\linewidth}
        \centering
        \includegraphics[width=\linewidth]{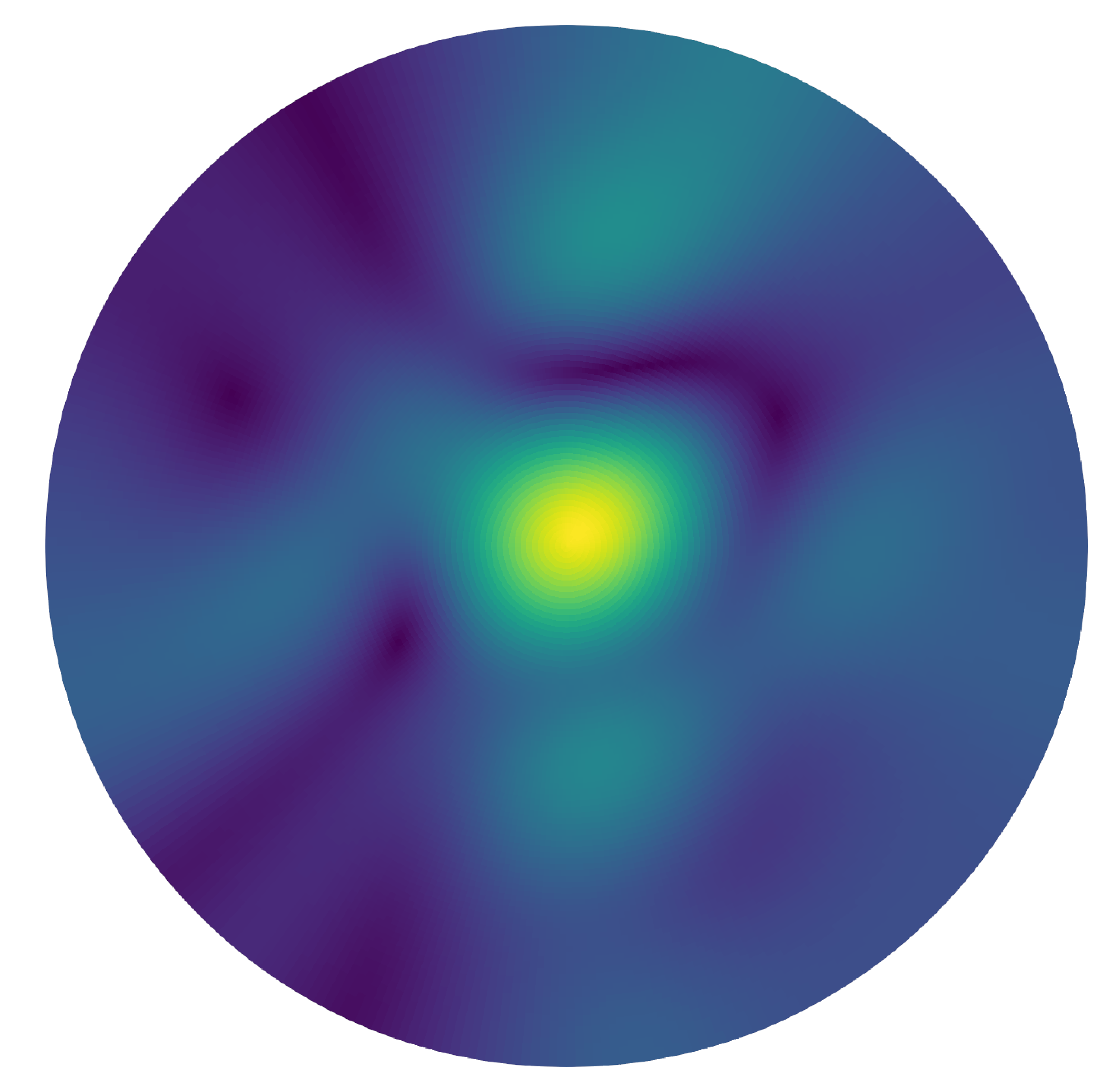}
        \caption{{Ground Truth}}
    \end{subfigure}
    \begin{subfigure}[t]{0.19\linewidth}
        \centering
        \includegraphics[width=\linewidth]{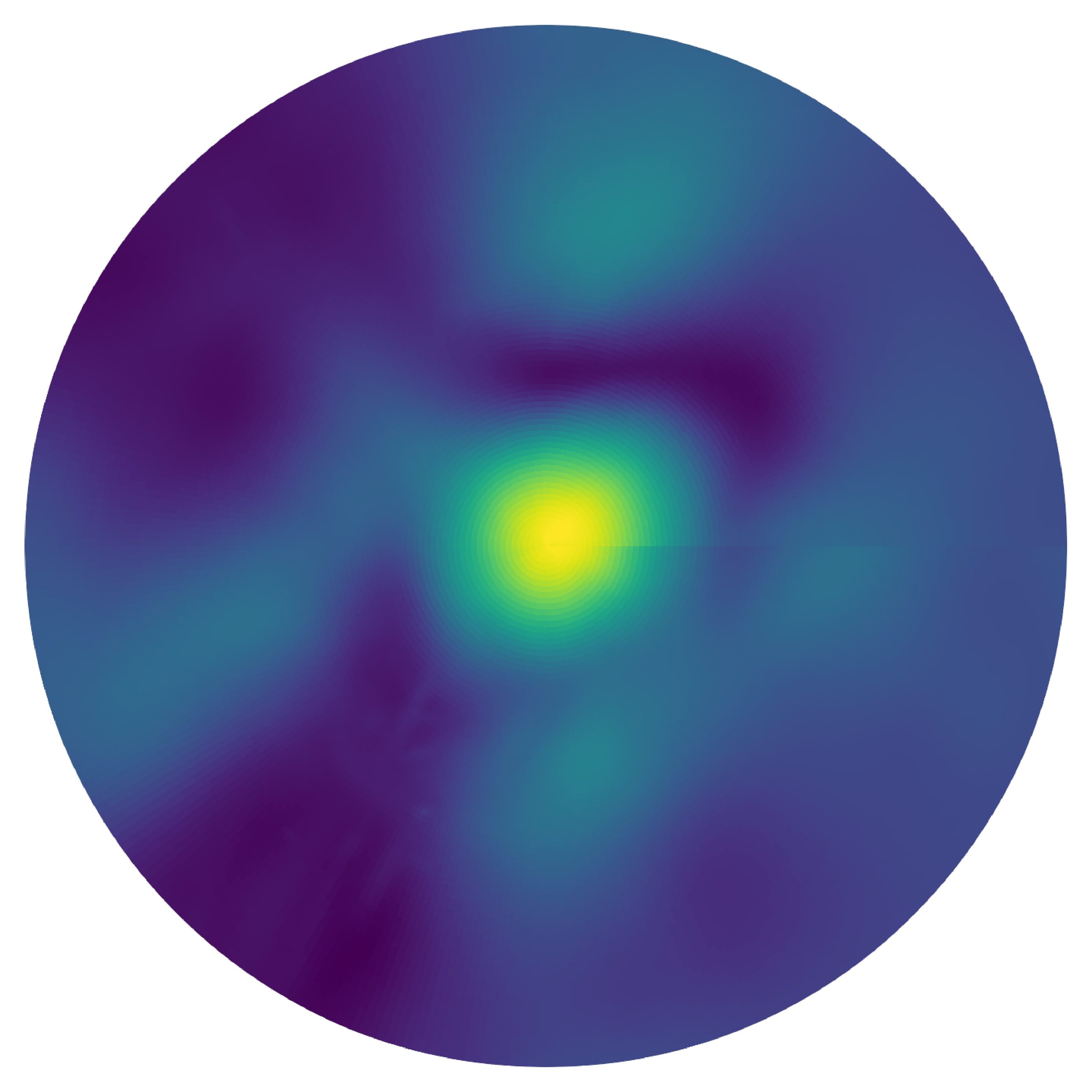}
        \caption{Proposed $\lambda = 0.01$}
    \end{subfigure}
    \begin{subfigure}[t]{0.19\linewidth}
        \centering
        \includegraphics[width=\linewidth]{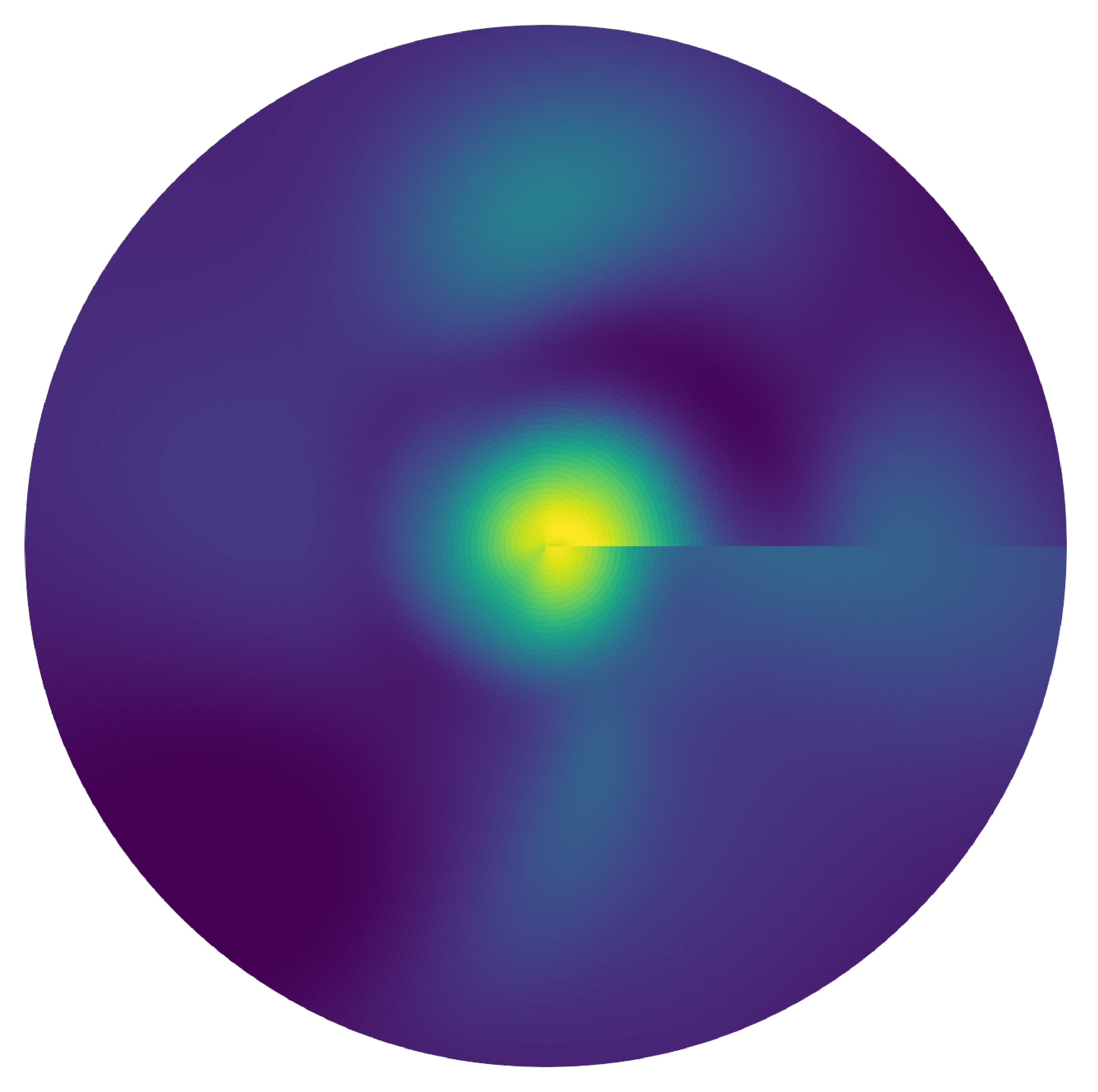}
        \caption{Proposed $\lambda = 0.03$}
    \end{subfigure}
    \begin{subfigure}[t]{0.19\linewidth}
        \centering
        \includegraphics[width=\linewidth]{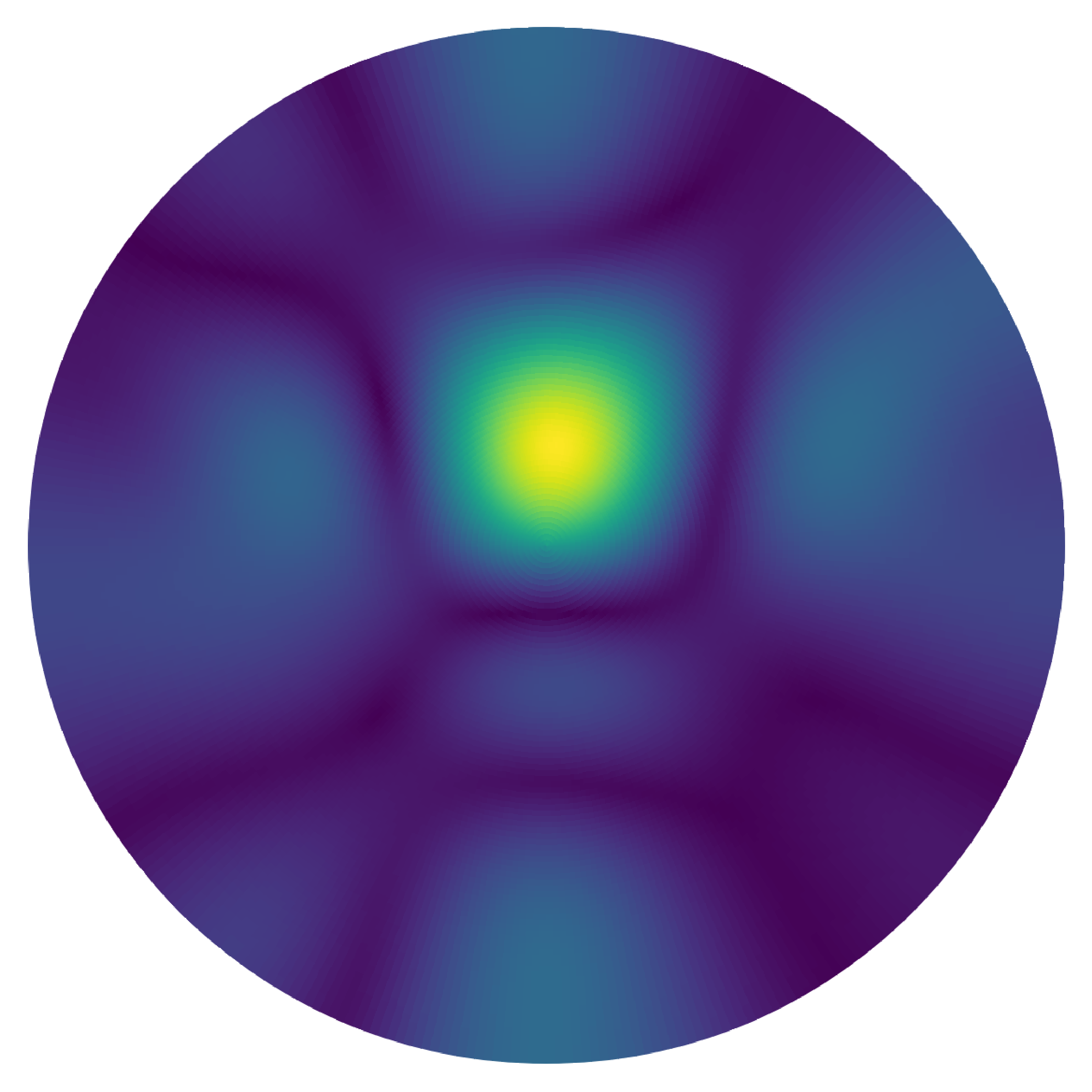}
        \caption{$\text{NeRF}^2$}
    \end{subfigure}
    \begin{subfigure}[t]{0.19\linewidth}
        \centering
        \includegraphics[width=\linewidth]{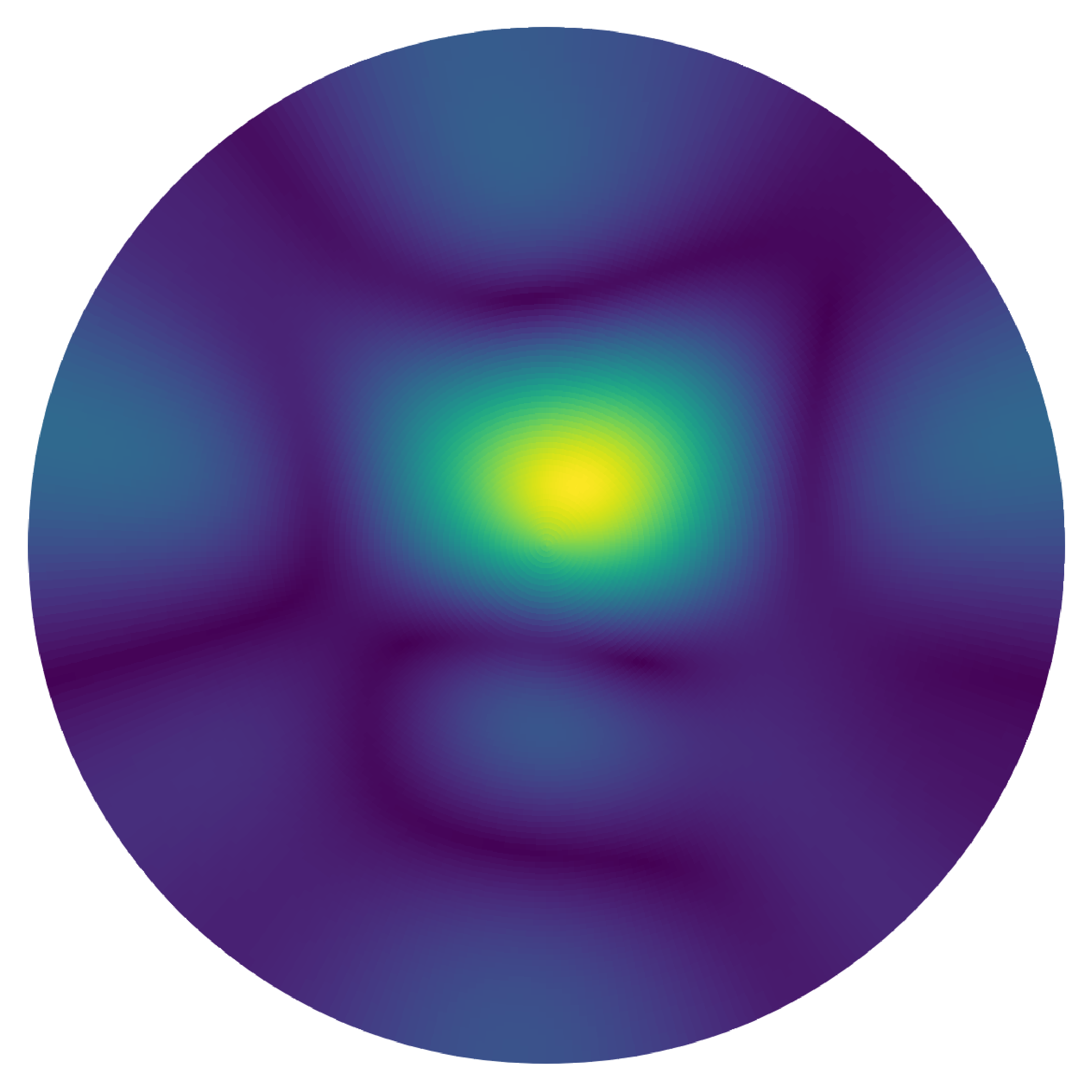}
        \caption{Ray-tracing}
    \end{subfigure}

    \vspace{-0.2cm}
    \caption{Comparison of spatial spectrum reconstruction under different methods.}
    \label{fig:spatial_spectrum}
\end{figure}

\subsection{Visualization}
Finally, we provide visualizations for a better understanding of the proposed {P-WRFGS} scheme. We first visualize the 3D Gaussian primitives of the {P-WRFGS} models trained with $\lambda = 0$ using different initialization methods. The geometry layout of the scene, i.e., $\mathcal{P}$, is also provided as a reference. Fig. \ref{fig:visualize_pc} shows that the 3D Gaussian primitives initialized from the point cloud remain better aligned with $\mathcal{P}$, whereas those from random initialization deviate substantially. This can be further confirmed by calculating the Chamfer distances between the Gaussian centers $\{\bm{\mu}_i\}_{i=1}^{N}$ and the original point cloud $\mathcal{P}$ for these two methods. The Chamfer distance is defined as:
\begin{align}
    { d_{\text{CD}} \triangleq \frac{1}{|\mathcal{P}|} \sum_{\bm{x}\in \mathcal{P}} \min_{i} \|\bm{x} - \bm{\mu}_i\|^2_2 + \frac{1}{N} \sum_{i=1}^{N} \min_{\bm{x}\in {\mathcal{P}}} \|\bm{\mu}_i - \bm{x}\|^2_2.}
    \label{eq:chamfer_dist}
\end{align}
Empirically, $d_{\text{CD}} = 0.22$ and $d_{\text{CD}} = 7.51$ for the model initialized using $\mathcal{P}$ and the random initialization, respectively.

{We further visualize the spatial spectrum achieved by {P-WRFGS} with $\lambda \in \{0.01, 0.03\}$. As shown in Fig. \ref{fig:spatial_spectrum}, {P-WRFGS} with $\lambda = 0.01$ ($N = 2613$) closely matches the ground-truth spectrum and outperforms both the $\text{NeRF}^2$ and ray-tracing baselines. However, the reconstruction of the model trained with $\lambda = 0.03$ significantly deviates from the ground truth, since $N = 24$ Gaussian primitives cannot provide sufficient information to reconstruct the spectrum and the prediction degrades visibly.}

\section{Conclusion}
We presented {P-WRFGS}, an efficient wireless radiance field modeling framework based on 3D Gaussian splatting with learnable masks. {By pruning less significant Gaussian primitives and jointly optimizing the rendering and regularization losses, {P-WRFGS} achieves up to 100$\times$ storage reduction and 7$\times$ rendering speedup with minimal quality loss.} 
%We also introduced a geometry-aware initialization using a 3D point cloud of the scene, which further improves the quality-efficiency trade-off. 
Numerical results verify that {P-WRFGS} is a practical, real-time alternative to $\text{NeRF}^2$ and ray-tracing for indoor channel modeling.

\appendices

\bibliographystyle{IEEEtran}
\bibliography{Hua_WCL2026-2977}

\end{document}